\documentclass[12pt]{iopart}
\usepackage{iopams}  
\usepackage{graphicx}
\usepackage{dcolumn}
\usepackage{bm}

\newcommand\op[1]{\mathop{\rm #1}\nolimits}
\newcommand\p{\partial}
\begin{document}

\title[]{A stochastic theory for temporal fluctuations in self-organized critical systems}

\author{M. Rypdal$^*$ and K. Rypdal$^\dag$  \\~ }

\address{$^*$Department of Mathematics and Statistics, University of Troms{\o}, Norway \\ $^\dag$Department of
Physics and Technology, University of Troms{\o}, Norway}
\ead{Martin.Rypdal@matnat.uit.no}
\begin{abstract}
A stochastic theory for the toppling activity in sandpile models is developed, based on a simple mean-field assumption about the toppling process. The theory describes the process as an anti-persistent  Gaussian walk, where the diffusion coefficient is proportional to the  activity.  It is formulated as a generalization of the It\^{o} stochastic differential equation with an anti-persistent fractional Gaussian noise source.  An essential element of the theory is re-scaling  to obtain a proper  thermodynamic  limit, and it captures all   temporal
features of the toppling process  obtained by numerical simulation of the Bak-Tang-Wiesenfeld sandpile in this limit.

\end{abstract}

\pacs{05.65.+b, 45.70.Ht, 02.50.Ey, 89.75.Da}

\maketitle
\section{Introduction}
The existence of  self-organized critical dynamics in complex systems has traditionally been  demonstrated
through numerical simulation of certain classes of cellular  automata  referred to as sandpile models
\cite{Jensenbook}. Non-linear, spatio-temporal dynamics is always essential for the emergence of SOC behavior, but the details
of this dynamics for a specific natural system is often poorly understood and/or not accessible to observation. In
many cases the information available is in the form of time-series of spatially averaged data like stock-price
indices, geomagnetic indices, or global temperature data. For scientists who deal with such data a natural
question to ask is: are there specific signatures of SOC dynamics that can be detected from such data?

In this letter we shall report some results which provide a partial answer to such a
question. Some important statistical features of the toppling activity are common to most weakly driven sandpile models described in the literature, and these are used to formulate a stochastic model for the toppling activity signal. A benchmark case  against which our results are tested, is a numerical study of the Bak-Tang-Wiesenfeld (BTW) sandpile \cite{BTW}. A crucial step in our work is a re-scaling of the dynamical variables which allows a natural passage to the thermodynamic (continuum) limit. We  demonstrate that this leads to  new results concerning SOC scaling laws.  We find that the probability density function (pdf) for the toppling activity is a stretched exponential or close to the Bramwell-Holdsworth-Pinton distribution \cite{Bramwell}, depending on whether the sandpile is so slowly driven that avalanches are well separated, or it is driven so hard that several avalanches run  simultaneously. The pdf for avalanche durations is unique in the thermodynamic limit, but  is not a power law, unless we redefine the meaning of an avalanche to be the activity burst  between successive times for which the activity rises above a positive threshold. Implementing such a threshold yields an exponent for the avalanche duration pdf of 1.63, in agreement with \cite{maya}, but in contradiction to \cite{Lubeck}. It also gives power-law quiet-time statistics as in \cite{maya} and thus refuting the claim in \cite{Bofetta} that SOC implies power-law distributed avalanche durations, but Poisson-distributed quiet times.

The sandpile models considered in this short paper deal with a
$d\geq 2$-dimensional lattice  of $N^d$ sites each of which are occupied by a certain integer number of
quanta which we conveniently can think of as sand grains. The  dynamics on the lattice is given by a toppling rule
which implies that if the number of grains on a site exceeds a prescribed threshold, the grains on that site are
distributed to its nearest neighbors. If the occupation number of some of these neighbors exceed the toppling
threshold these sites will topple in the next time step, and the dynamics continues as an avalanche until all
sites are stable. The details of this toppling rule can vary, but a useful theory for a broad class of natural phenomena should not be very sensitive to such detail.

In natural systems the SOC dynamics is usually driven by some weak random external forcing. In sandpile models
this can be modeled by   dropping of sand grains at randomly selected sites at  widely separated times. In
numerical algorithms this is often done by dropping sand grains only at those times when no avalanche is running.
This ensures that the drive does not interfere with the avalanching process. Usually it will  then only take a few
time steps from one avalanche has stopped until a new starts, so for a large system the quiet times between avalanches will appear insignificant compared to their durations. 

A more physical drive would be to drop
sand also during avalanches. If the dropping  rate is slower than the typical duration of a system-size avalanche the drive would still not interfere
with the avalanche dynamics, but the quiet times would depend on the statistical distribution of dropping times, which is typically a Poisson distribution. In many natural systems, however, avalanching occurs all the time, corresponding to a higher driving rate. In such cases, and also
because  there will always be noise in time-series data,  we cannot identify the start and
termination of an avalanche from a zero condition of our observable. In practice we have to define avalanches as
{\em bursts} in the time series identified by a threshold on the signal \cite{maya}. In a sandpile simulation such
 bursts are correlated and therefore the quiet times between the
bursts are power-law distributed even if the dropping of sand grains is chosen to be a Poisson process. Hence if 
focus is on modeling features that can be detected in observational data we  shall think of
avalanches as activity bursts  starting and terminating at a non-zero threshold value. Moreover, one of the main results of this work is that power-law shape of the pdf for avalanche duration  is true  {\em only} if one defines avalanches in this way. 

\section{The stochastic model}
\noindent
\begin{figure}[t]
\begin{center}
\includegraphics[width=8.5cm]{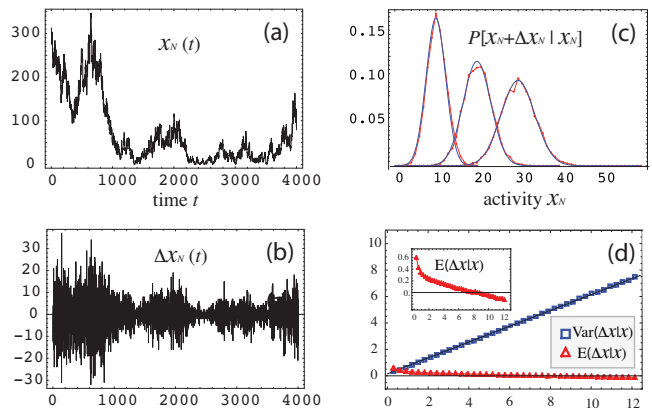}
\caption{a): A realization of the toppling activity $x_N(t)$ in the BTW sandpile. b): The increments  $\Delta x_N(t)=x_N(t+1)-x_N(t)$ of the trace in (a), showing that $\Delta x_N(t)$ is large when $x_N(t)$ is large. c): Conditional pdfs of $x_N+\Delta x_N$ for $x_N=10,\, 20,\, 30$ respectively. d) The conditional mean and variance of $\Delta x_N$ versus $x_N$.} \label{Fig1}
\end{center}
\end{figure}
We shall assume that the lattice has linear extent $L=1$ with $N^d$ sites, so the thermodynamic limit $ N\rightarrow \infty$ can be thought of as a continuum limit.  The
sandpile evolves in discrete time steps labeled by $k=1,2,3\ldots$, and the  number of sites whose occupation
number exceeds the toppling threshold at time $k$  is called the toppling activity  $x_N(k)$. The toppling
increment is $\Delta x_N(k) \stackrel{\mathrm{def}}{=} x_N(k+1)-x_N(k)$. Let us define two active sites as
dynamically connected if they have at least one common  nearest neighbor, and   define a connected cluster as a
collection of active sites which are linked trough such connections. From numerical simulations of sandpiles we
observe that such clusters never consist of more than a few elements and that the instantaneous number of clusters $n_N$ increases
in proportion to $x_N$. This  implies that at each time $k$ we can label the  clusters by $i=1,\dots,c
x_N(k)$, where $c<1$ is a constant depending on the specific toppling rule and the dimension $d$ of the sandpile.
We can then  decompose the increment $\Delta x_N(k)$ into a sum of  local increment contributions
$\xi_{N,i}(k)$ produced by each of the  clusters, i.e.
$
\Delta x_N (k)=\sum_{i=1}^{c x_N(k)} \xi_{N,i}(k).
$
We think of the local increment contributions as random variables which take values in a finite sample space. Indeed, if each cluster $i$ only consists of a single overcritical site, then $\xi_{i,N}$ takes values in the set $\{-1,0,\dots,2d-1\}$. 

As a first step to a stochastic model we make a mean-field assumption \cite{Tang,Ivashkevich}, which impiles that $\xi_{N,i}(k)$ and $\xi_{N,j}(k)$ are statistically
independent  for $i \neq j$. Then the central limit theorem  states
that in the limit $N\rightarrow \infty$, $x_N(k)\rightarrow \infty$ the conditional probability density  $P[\Delta x_N(k) | x_N(k)]$ of an increment $\Delta x_N(k)$,
given $x_N(k)$, is Gaussian with variance $\sigma^2\,
x_N(k)$, where  $\sigma^2=c^2(E[\xi_{N,i}^2|x_N] -(E[\xi_{N,i}|x_N])^2)$. 
This has been verified numerically in the two-dimensional BTW-model as  shown in Fig.\ \ref{Fig1}. The figure demonstrates the need to introduce a  conditional probability: The conditional variance of the increments is proportional to $x_N$ and the conditional mean 
is not zero.
 
In fact, numerical simulations show that the the conditional mean increment,  $E[\Delta x_N|x_N]$, is positive for small $x_N$, reflecting the natural tendency for the activity to grow when it is small. On the other hand the mean increment  decays exponentially to zero for moderate $x_N$, and becomes negative when $x_N$ is comparable to the activity of a system-size avalanche, reflecting the limiting influence of the finite system size. These effects will be incorporated as a drift-term correction to the model, but for now we consider for simplicity of argument a Gaussian process  with non-stationary increments and no drift term:
\begin{equation}
\Delta x_N(k)=\sigma\, \sqrt{x_N(k)}\,w(k)\,, \label{gaussincrements2}
\end{equation}
where $w(k)$ is a stationary Gaussian 
stochastic process with unit variance. From the numerical sandpile data (see Fig.\ \ref{Fig1}) we observe that the {\em normalized}  toppling process
\begin{equation*}
W(k)\stackrel{\mathrm{def}}{=}\sum_{k'=0}^k w(k')=\sum_{k'=0}^k\,\frac{\Delta x_N(k')}{\sigma \sqrt{x_N(k')}}
\end{equation*}
has the characteristics of a fractional Brownian walk with Hurst exponent $H\approx 0.37$ on time scales shorter than the characteristic growth time for a system-size avalanche, consistent with a power spectrum which scales like $f^{-1.74}$. Thus we  model the normalized increment
process as $w(k)=W_H(k+1)-W_H(k)$, where $W_H(k)$ is a fractional Brownian walk with Hurst exponent $H$. For the transition to the thermodynamic
limit, where time will become a continuous variable, we can think about $W_H(k)$ as the result of a discrete
sampling of the (continuous-time) fractional Brownian motion (fBm) $W_H(t)$. This process has the property
$\langle |W_H(t+\tau)-W_H(t)|^2\rangle =\tau^{2H}$. We now have a stochastic difference equation
\begin{equation}
\Delta x_N(k)=\sigma\,\sqrt{x_N(k)}\,(W_H(k+1)-W_H(k)). \label{discrstocheq}
\end{equation}
Numerical simulations show that $x_N \sim N^{D_1}$ 
\footnote{The BTW model does not exhibit perfect finite-size scaling \cite{Christensenbook} and hence the scaling $x_N \sim N^{D_1}$ is not valid for very large activity. The effect of imperfect scaling with increasing $N$ can be built into Eq.~\ref{SDE} through an $N$-dependent drift term. However, the  distributions of duration and size of  sub-system size avalanches (defined by a threshold $X_c>0$)  is not sensitive to this feature of the BWT model. We have given a  detailed treatment of this problem in \cite{RypdalPRE}.}, 
where $0<D_1\leq d$ can be interpreted as a fractal dimension
of the set of active sites imbedded in the $d$-dimensional lattice space. This property is used to re-scale
$x_N(k)$  such  that it has  a well-defined limit as $N \to \infty$. We also have to re-scale the time variable
by letting $t=k \Delta t$, where $\Delta t =N^{-D_2}$. The value of $D_2$ will become apparent  if we define the normalized activity variable
$X_N(t)=N^{-D_1}x_N(t /\Delta t)$, such that the corresponding increment becomes
\begin{equation}
\Delta X_N(t)=N^{HD_2-D_1}\,\sigma\,\sqrt{X_N(t)}\,\Delta W_H(t)\,, \label{stocheq1}
\end{equation}
where $\Delta W_H(t)=W_H(t+\Delta t)-W_H(t)$.
A well-defined thermodynamic limit $N\rightarrow \infty$ requires
$D_2=D_1/2 H$, for which Eq.\ (\ref{stocheq1}), by introduction of the limit function $X(t)=\lim_{N \to
\infty} X_N(t)$, reduces to the stochastic differential equation
\begin{equation} \label{SDE}
dX(t)=f(X)\,dt+\sigma \,\sqrt{X(t)}\, dW_H(t),
\end{equation}
where we have  heuristically added a drift term $f(X)\,dt$ to account for the non-zero mean of the conditional increment. We take $f(X)$ to be an exponentially decaying function based on the numerical results from the sandpile.
In the $2$-dimensional BTW model we find that $D_1 \approx 0.86$ and hence $D_2=1.16$. This defines re-scaled coordinates $X_N=x_N/N^{0.86}$ and $t_N=k/N^{1.16}$.

\section{Analysis of avalanches}
\noindent
\begin{figure}[t]
\begin{center}
\includegraphics[width=8.5cm]{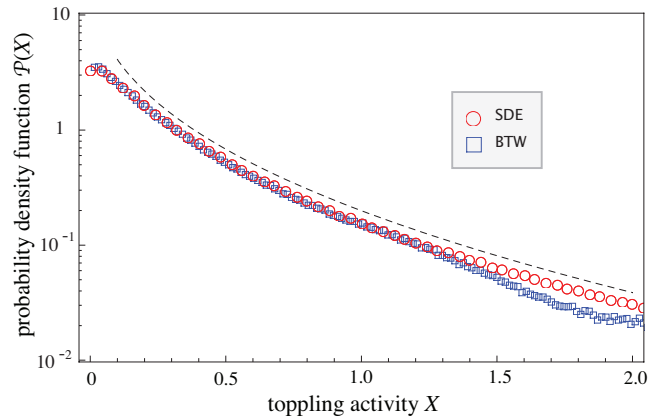}
\caption{Logarithmic plots of ${\cal P}(X)$ from simulations of the $2$-dimensional BTW sandpile for $N=1024$. Also shown is ${\cal P}(X)$ found from simulations of Eq.\ (\ref{SDE}), and a stretched exponential fit (dashed curve, vertically shifted for visibility). All pdfs are scaled to unit variance.} \label{Fig2}
\end{center}
\end{figure}
A time series $X(t)\geq 0$, representing a succession of avalanches with zero quiet times, can be constructed numerically from   the discrete-time version of Eq.\ (\ref{SDE}) by integrating the equation using realizations of the fractional Gaussian noise process $\Delta W_H(t)$. At those times when $X(t)$ drops below zero we consider  the avalanche as terminated, and a new, independent realization of  $\Delta W_H(t)$ is generated and used to produce the next avalanche.  From long, stationary  time-series generated from the stochastic model and from the  sandpile model this way, we can construct  pdfs ${\cal P}(X)$ which turn out to give almost identical results for the two models (see Fig.\ \ref{Fig2}). The shape of this pdf  is universal in the thermodynamic limit:  a stretched exponential ${\cal P}(X)\sim \exp{(-aX^{\mu})}$ with $\mu \approx  0.5$. A different pdf appears if  the time-series are constructed by launching the avalanches at random times (Poisson-distributed) with characteristic time between launches shorter than the growth time of a system size avalanche. In this case several avalanches may run simultaneously, and ${\cal P}(X)$ from both models are close to the  Bramwell-Holdsworth-Pinton distribution, which was claimed to be valid for the toppling-activity in the BTW-model in \cite{Bramwell}.

\begin{figure}[t]
\begin{center}
\includegraphics[width=8.5cm]{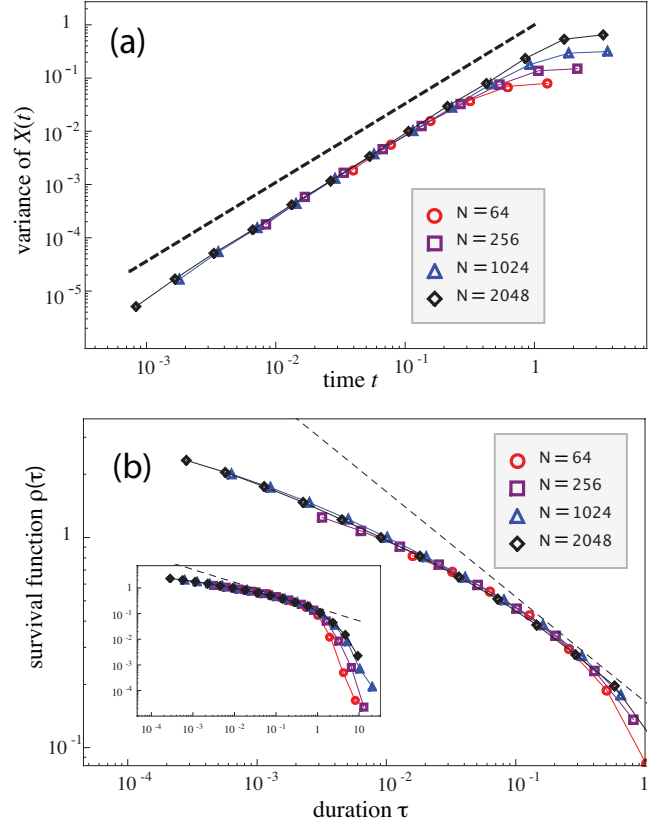}
\caption{a) Double-logarithmic plots of the variance of $X(t)$ with respect to the pdf $P(X,t)$. The variance grows like $t^{2h}$, with $h=2H=0.74$ for times less than the duration of a system size avalanche. b) Double-logarithmic plots of the survival function $\rho(\tau)$ in the re-scaled coordinates $X_N$ and $t_N$, demonstrating that the pdf of avalanche durations is not a power-law. The dotted line has slope $-0.5$.} \label{Fig3}
\end{center}
\end{figure}

Consider a solution of Eq.\ (\ref{SDE})   with initial condition $X(0)=Y>0$, and let $P(X,t)$ be the evolution of the density distribution in $X$-space of an ensemble of realizations of the
stochastic process $X(t)$ all launched at activity $X=Y$ at time $t=0$. Every realization $X(t)$ will  sooner or later terminate at a
finite time $t=\tau$ for which $X(\tau-1)>0$ and $X(\tau)\leq 0$, and then  we remove it from the ensemble.  $P(X,t)$ contains information about  all   commonly considered avalanche characteristics. For example, it is easily found from from Eq.\ (\ref{SDE}) that, on time scales  shorter than the growth time of a system-size avalanche, $X(t)$ is a self-similar process with non-stationary increments and self-similarity exponent $h=2H$ \cite{RypdalPRE}. Hence the variance of $X(t)$ with respect to $P(X,t)$ will scale as $\sim t^{2 h}$. That this relation holds for the 2-dimensional BTW model can easily be verified through numerical simulation (Fig.\ \ref{Fig3}(a)).   

We can also compute the survival probability $\rho(\tau)=\int_0^\infty P(X,\tau)\,dX$, which is the probability that a realization of an avalanche has not terminated at the time $\tau$. This function is related to the pdf for avalanche durations by $p_{\op{dur}}(\tau)=-\rho'(\tau)$, so that $p_{\op{dur}}(\tau)$ is a power law if and only if $\rho(\tau)$ is a power law. Fig.\ \ref{Fig3}(b) shows the function $\rho(\tau)$ for numerical simulations of the BTW sandpile in the re-scaled coordinates $X_N$ and $t_N$,  demonstrating that the pdf for avalanche durations does {\em not} represent a power law. The power-law form $\rho(\tau) \sim \tau^{0.5}$ proposed in \cite{Lubeck} can only be obtained as a tangent to the log-log plot of $\rho(\tau)$ at a given duration time $\tau$, and the slope of this tangent depends crucially on the duration time $\tau$ for which this tangent is drawn. 

The situation changes if we let all avalanches terminate when $X$ drops below a small threshold $X_c>0$ as proposed in \cite{maya}. In this case avalanche durations are the return times to the line $X=X_c$, and by changing coordinates to $Y=X-X_c$ we see that this corresponds to the return times to the time axis of the process given by the stochastic differential equation
$dY(t)=\sigma\,\sqrt{X_c+Y(t)}\,dW_H(t)$.   
For small avalanches where $X(t)-X_c \ll X_c$ we can approximate this expression with $dY(t)=\sigma\,\sqrt{X_c}\,dW_H(t)$, i.e. can approximate $Y(t)$ by a fractional Brownian motion with Hurst exponent $H$. Using the result of Ding and Yang \cite{DingYang} on the return times of a fractional Brownian motion we get $p_{\op{dur}}(\tau) \sim \tau^{2-H}=\tau^{-1.63}$. 

\begin{figure}[h!]
\begin{center}
\includegraphics[width=8.5cm]{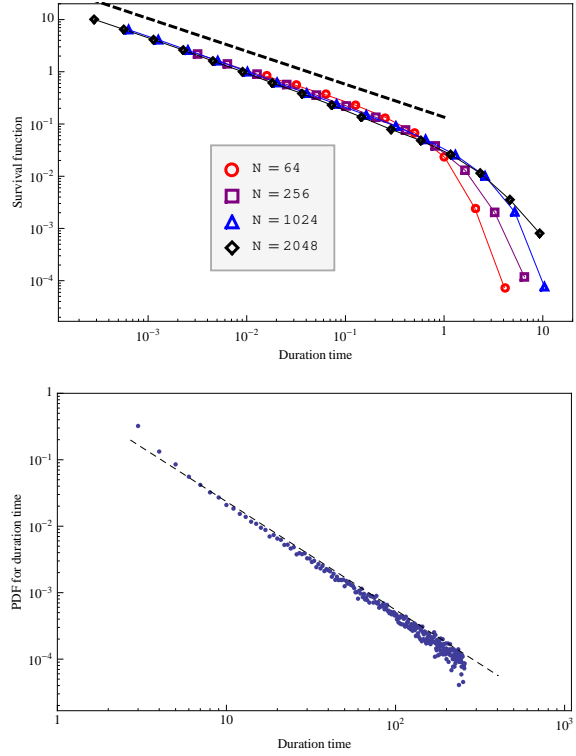}
\caption{a) The survival function for the BTW sandpile in re-scaled coordinates $X_N$ and $t_N$ for $N=64,256,1024,2048$ when the durations are defined by putting a small threshold $X_c$ on the toppling activity. The function shows power-law behavior with exponent $-0.63$ for avalanches smaller than system size. b) The pdf for duration times from simulations of Eq.\ (\ref{SDE}) when avalanches are defined in the same way as for the sandpiles. The dotted line has slope $-1.63$.} \label{Fig4}
\end{center}
\end{figure}

Numerical simulations of the BTW model verify this result: The survival function $\rho(\tau)$ becomes a power law on time scales shorter than a system-size avalanche (see Fig. \ref{Fig4}(a)), and the slope of the graph in a log-log plot is approximately $-0.63$, which corresponds to a scaling of the pdf for duration times on the form $p_{\op{dur}}(\tau) \sim \tau^{-1.63}$. 
The result is also reproduced by simulations of Eq.\ (\ref{SDE}) with an exponentially decaying drift term. Fig.\ \ref{Fig4}(b) shows the log-log plot of the pdf for duration times in the stochastic differential equation  and  a line with slope $-1.63$, demonstrating that  the avalanche statistics in the BTW sandpiles is captured by the stochastic differential equation. 
\noindent
\noindent
From the scaling $\rho(\tau) \sim \tau^{-\alpha}$ we can deduce an exponent for the pdf of avalanche size as well. 
On the time scales where the toppling activity can be approximated by a fractional Brownian motion $W_H(t)$, the signal disperses with time as $X \sim t^{H}$, the size of an avalanche of duration $\tau$ scales like $S(\tau) \sim \int_0^\tau t^H\,dt \sim \tau^{H+1}$. Assuming that the pdf for avalanche size is on the form $p_{\op{size}}(S) \sim S^{-\nu}$, the relation $p_{\op{size}}(S)\,dS=p_{\op{dur}}(\tau)\,d\tau$ yields 
$\tau^{-\nu(H+1)+H} \sim \tau^{-\alpha-1}$, so 
\begin{equation} \label{rel}
\nu=\frac{H+\alpha+1}{H+1}=\frac{2}{H+1}\,.
\end{equation}
With $H=0.37$ we obtain $\nu=1.46$. 

We also remark that if we omit the drift term and let $H=1/2$ and $X_c=0$ we obtain the so-called mean-field theory of sandpiles. In this case the stochastic differential equation has a corresponding Fokker-Planck equation 
\begin{equation*}
\frac{\p P}{\p t}=\frac{\sigma^2}{2}\,\frac{\p^2}{\p X^2}{\left( X P \right)}\,.
\end{equation*}
If we solve this equation on the interval $[0, \infty)$ with absorbing boundary conditions in $X=0$ we can obtain an analytical expression for  $P(X,t)$, and from some straightforward algebra we find for large $\tau$ that
$
p_{\op{dur}}(\tau)\sim \tau^{-2}
$ \cite{RypdalPRE}.
Since $X_c=0$ we can not approximate the toppling activity by a Brownian motion on any scale and thus $X(t)$ diverges like $\sim t^h$, where $h=2H$. 
By replacing $H$ with $h=2H$ in Eq.\ (\ref{rel}) we get $p_{\op{size}}(S) \sim S^{-3/2}$, in agreement with previous mean field approaches \cite{Tang,Ivashkevich}. 

\section{Concluding remarks}
We point out that the validity of Eq.~\ref{SDE} is not restricted to the BTW model. For instance, the equation has been verified for the Zhang model \cite{RypdalPRE,Zhang}, though with a different Hurst exponent $H$.  Time series of global quantities derived from numerical simulation of  different sandpile and turbulent fluid systems can be shown to be adequately described by Eq.~ 4, where $H$ and the specific form of the drift term depend on the system at hand \cite{RypdalPRE}.

\newpage

\end{document}